\documentclass[aps,physrev,twocolumn,superscriptaddress,nofootinbib]{revtex4-2}

\usepackage{xcolor}
\usepackage{graphicx}
\usepackage{booktabs}
\usepackage{amsmath}

\newcommand{\Fstat}{$\mathcal{F}$-statistic}
\newcommand{\Lcal}{\mathcal{L}}
\newcommand{\Lbar}{\bar{\mathcal{L}}}
\newcommand{\Lthres}{\Lbar_\mathrm{th}}
\newcommand{\Ltail}{\Lbar_\mathrm{tail}}
\newcommand{\pfa}{p_\mathrm{fa}}
\newcommand{\pcorr}{\pfa^\mathrm{corr}}
\newcommand{\rmd}{\mathrm{d}}
\newcommand{\aproj}{a_\mathrm{p}}
\newcommand{\Pbin}{P_\mathrm{b}}
\newcommand{\hsens}{h_0^{95\%}}
\newcommand{\depth}{\mathcal{D}^{95\%}}

\begin{document}

\title{Spotlight searches for continuous gravitational waves triggered on radiometer candidates in LIGO O4a data}

\newcommand{\umich}{\affiliation{Department of Physics, University of Michigan, Ann Arbor, MI 48109, USA}}
\newcommand{\anu}{\affiliation{OzGrav-ANU, Centre for Gravitational Astrophysics, Research School of Physics and Research School of Astronomy \& Astrophysics, The Australian National University, ACT 2601, Australia}}

\author{Alan M. Knee}\email[]{aknee@umich.edu}\umich
\author{Keith Riles}\umich
\author{Ling Sun}\anu

\date{\today}

\begin{abstract}
We report on a follow-up search for continuous gravitational waves (CWs) targeting sub-threshold candidates from the O4a LIGO--Virgo--KAGRA (LVK) all-sky all-frequency (ASAF) directed radiometer analysis. 
We analyze a total of $562$ ASAF candidates using Advanced LIGO data from the first part of the fourth LVK observing run. 
Each candidate comprises a narrow $1/32$~Hz frequency band paired with a ${\sim}13$~deg$^2$ sky pixel, defining the search parameter space for our analysis.
Our search pipeline leverages the \Fstat{} matched filter equipped with a hidden Markov model to enable tracking of a potentially frequency-wandering CW signal. 
Of the 562 initial candidates, we obtain $21$ outliers with false alarm probability less than $5\%$ that survive all instrumental vetoes. 
None of the outliers are recovered at comparable statistical significance in the second part of the fourth observing run, from which we conclude that no convincing CW signals are detected.
We estimate the sensitivity of our analysis by recovering simulated signals added to real detector data.
Across the $20\text{--}1726$~Hz frequency band, our search achieves $95\%$ detection efficiency for strain amplitudes in the range of $h_0\sim (0.63\text{--}6.3)\times10^{-25}$ with respect to isolated neutron stars (NSs), and $h_0\sim (0.82\text{--}7.1)\times10^{-25}$ with respect to NSs in long-period ($>1$~yr) binary systems.
The large range in sensitivity is due to the strong frequency dependence of detector noise, particularly at low frequencies.
\end{abstract}


\maketitle

\section{Introduction\label{sec:intro}}
The search for persistent, quasi-monochromatic gravitational waves (GWs), known as continuous waves (CWs), is a primary science objective of the Advanced Laser Interferometer Gravitational-Wave Observatory (LIGO)~\cite{LIGOScientific:2014pky}, Advanced Virgo~\cite{VIRGO:2014yos}, and KAGRA~\cite{KAGRA:2018plz} ground-based detectors.
In their sensitive $10\text{--}2000$~Hz frequency band, the principal CW sources are expected to be rapidly rotating galactic neutron stars (NSs) that sustain long-lived non-axisymmetric deformations~\cite{Lasky:2015uia, Sieniawska:2019hmd, Haskell:2021ljd, Riles:2022wwz, Wette:2023dom}, either from rigid crust mountains~\cite{Gittins:2024zbg} or unstable $r$-mode oscillations~\cite{Haskell:2015iia}.
The typical strain amplitude is predicted to be extremely weak, on the order of $h\lesssim 10^{-24}$, and CWs have thus far eluded detection.
The discovery of a CW signal, especially in conjunction with electromagnetic observations of the same source, would provide a unique glimpse into extreme NS interiors.
Searches for CWs with GW detectors have been carried out for decades, setting progressively tighter constraints on NS ellipticities for a diverse range of celestial targets~\cite{
KAGRA:2022dwb,
LIGOScientific:2020qhb,
Steltner:2023cfk, 
McGloughlin:2025iyx, 
McGloughlin:2025eso, 
LIGOScientific:2026plm,
LIGOScientific:2025ouy, 
LIGOScientific:2021mwx,
KAGRA:2026sed, 
Ming:2025ehy,
LIGOScientific:2021hvc,
LIGOScientific:2021quq,
LIGOScientific:2021yby,
LIGOScientific:2025kei, 
LIGOScientific:2026qsb,
KAGRA:2025qeh, 
LIGOScientific:2021ozr, 
Middleton:2020skz, 
KAGRA:2022dqk,
LIGOScientific:2022enz,
Whelan:2023bha, 
KAGRA:2022osp,
Dunn:2025ttx,
Cheung:2026viy,
LIGOScientific:2021rnv, 
LIGOScientific:2025csr}.


CW search methods typically assume a coherent signal undergoing secular phase evolution.
While this is expected to be valid for most ``well-behaved'' NSs, there are certain scenarios where the assumption of full coherence breaks down.
For instance, it is well-known from radio observations that pulsars exhibit stochastic timing noise~\cite{2010MNRAS.402.1027H, 2010ApJ...725.1607S, 2012MNRAS.426.2507P, Parthasarathy:2019txt, Lower:2020mjq}.
This timing noise may be due to intrinsic processes or external forces acting on the star, such as in actively accreting binaries where accretion rate variability drives NS spin-wandering~\cite{Mukherjee:2017qme}.
A number of pulsars, including high-value CW targets such as Crab~\cite{Lyne:2014qqa, Keitel:2019zhb} and Vela~\cite{Espinoza:2020pow, KAGRA:2025qeh}, can also experience sudden and recurring phase discontinuities in the form of spin-up ``glitches''~\cite{Fuentes:2017bjx}.
Glitching pulsars are especially problematic for CW searches that lack electromagnetic input, as the size and times of any glitches would be unknown.
NSs residing in binary systems~\cite{Manchester:2004bp} pose another challenge for CW searches.
Though not a stochastic process, the orbital motion of the NS is an additional source of Doppler modulation, which is challenging to account for in a templated search due to the steep computational cost of the larger parameter space~\cite{Messenger:2011rg, Leaci:2015bka}.
Given these circumstances, it is valuable to explore alternative search techniques that are more robust to deviations from the canonical phase model of an isolated spinning NS.

One approach to detecting CWs with irregular phase evolution is to follow up candidates from stochastic GW background (SGWB) searches, in particular, the LIGO--Virgo--KAGRA (LVK) all-sky all-frequency (ASAF) directional analysis~\cite{KAGRA:2021rmt, LIGOScientific:2025bkz}.
Rather than filtering on pre-determined waveform templates, the ASAF analysis cross-correlates strain data from pairs of geographically separated detectors (``baselines'') in the LVK network, functioning as a GW radiometer search~\cite{Ballmer:2005uw, Mitra:2007mc, 2009PhRvD..80l2002T}.
Improvements to the ASAF technique since the O1-O2 era have greatly expanded its scope, enabling radiometric searches over the entire sky and in every frequency bin for unknown narrowband GW sources~\cite{Ain:2015lea, Thrane:2015aua, Goncharov:2018ufi, Ain:2018zvo, Suresh:2020khz}.
The modern ASAF method synthesizes the earlier narrowband and broadband radiometer methods, which could only perform frequency-resolved searches at specific sky directions~\cite{LIGOScientific:2016nwa, LIGOScientific:2019gaw, KAGRA:2021mth} or all-sky searches averaged over frequency, respectively.
ASAF analyses conducted to date~\cite{KAGRA:2021rmt, LIGOScientific:2025bkz} have searched frequencies between $20\text{--}1726$~Hz using a frequency resolution of $1/32$~Hz and a \textsc{HEALPix}\footnote{https://healpix.sourceforge.io}-based~\cite{Gorski:2004by, Zonca2019} sky grid consisting of $3072$ equal-area pixels.
This configuration gives ${\sim}10^8$ unique frequency-pixel pairs for establishing upper limits and identifying candidate signals of interest.

A crucial advantage of the ASAF framework is its model-agnosticism.
Since it is based on measuring correlated excess power in coarse-grained frequency bins, there is no requirement that the signal obey strict phase-coherence with respect to an intrinsic waveform model.
This flexibility potentially allows the ASAF method to detect quasi-monochromatic signals with irregular phase evolution that templated search methods are generally insensitive to.
Candidate frequency-pixels identified by the ASAF analysis are therefore of considerable interest to CW searches~\cite{KAGRA:2021rmt, LIGOScientific:2025bkz}, as deeper follow up of these candidates could reveal a previously unidentified CW source.

In this work, we systematically follow up $562$ sub-threshold candidates reported in the O4a LVK ASAF analysis~\cite{LIGOScientific:2025bkz} in search of evidence for CW emission.
The O4a ASAF analysis extends the preceding O3 analysis~\cite{KAGRA:2021rmt} by combining LIGO and Virgo data from the O1-O3 era through the first eight months of the fourth LVK observing run (O4a).
We employ a hidden Markov model (HMM)-based CW search pipeline to follow up these candidates, which was first prototyped in Ref.~\cite{Knee:2023toa} to follow up candidates from the O3 ASAF analysis~\cite{KAGRA:2021rmt}.
The HMM algorithm can accommodate a signal that wanders in frequency over time, providing robustness against stochastic spin-wandering, spin-up glitches, and other forms of poorly constrained phase evolution.
As such, HMM techniques have featured prominently in searches targeting accreting NSs~\cite{Middleton:2020skz, LIGOScientific:2021ozr, KAGRA:2022dqk, Vargas:2023gvd}, young supernova remnants~\cite{LIGOScientific:2021mwx, KAGRA:2026sed}, exotic boson clouds~\cite{LIGOScientific:2025csr}, and more recently, as a tool for stochastic candidate follow-up~\cite{Knee:2023toa}.
A dual-harmonic version of this technique was also used to search for long-transient CWs after a glitch in the Vela pulsar~\cite{KAGRA:2025qeh}.

We perform our analysis on publicly available Advanced LIGO data from O4a~\cite{LIGOScientific:2025snk}.
We also analyze data from the subsequent part of the fourth observing run (O4b)~\cite{LIGOScientific:2026jgl} for ASAF candidates found to be statistically significant in the initial O4a analysis.
We find no evidence in support of a detection, and estimate the sensitivity of our search empirically by recovering simulated signals injected into O4a data.

This paper is organized as follows: 
Sec.~\ref{sec:candidates} and \ref{sec:data} discuss the O4a ASAF candidates and the data set used for our follow-up analysis; 
in Sec.~\ref{sec:method}, we provide an overview of the search pipeline and the configuration used in this work; 
in Sec.~\ref{sec:results}, we present the results of our follow ups, post-processing of significant outliers, and estimate the sensitivity of our search; 
finally, we conclude in Sec.~\ref{sec:conclusions}.

\section{Candidates\label{sec:candidates}}

\begin{figure}[t]
\includegraphics[width=0.45\textwidth]{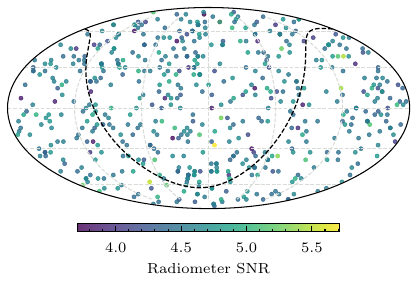}
\caption{Sky positions in equatorial coordinates and SNRs of the $562$ sub-threshold candidates identified in the O4a LVK ASAF analysis~\cite{LIGOScientific:2025bkz}. Right ascension starts at the right edge of the map and increases towards the left. The dashed line shows the Galactic plane.
\label{fig:candidates}}
\end{figure}

The $562$ O4a ASAF candidates are shown in Fig.~\ref{fig:candidates}.
The O4a ASAF analysis adopted the same frequency and angular resolutions used in the O3 analysis~\cite{KAGRA:2021rmt}, with each candidate corresponding to a $1/32$~Hz frequency bin paired with a four-sided curvilinear sky pixel of area ${\sim} 13.4$~deg$^2$.
The candidate frequencies range from $23.8125$~Hz up to $1724.5$~Hz.
All candidates were found to be of sub-threshold significance, where the ASAF analysis defines ``sub-threshold'' as being in the 99th signal-to-noise ratio (SNR) percentile within a neighborhood of nearby frequency bins, but falling below a global 95th percentile threshold calculated against the entire dataset~\cite{LIGOScientific:2025bkz}.
The number of sub-threshold candidates is comparable to the O3 analysis~\cite{KAGRA:2021rmt}, and the candidates appear uniformly distributed in frequency and isotropic over the sky. 
Furthermore, the O4a analysis obtained a largely disjoint set of candidates from the O3 analysis; only $23$ candidates are common to both the O3 and O4a sets of sub-threshold candidates.
Of these $23$ common candidates, only one yielded an outlier in the O3 follow-up~\cite{Knee:2023toa}, which was vetoed as a detector noise artifact.

\section{Dataset\label{sec:data}}

\begin{figure*}[t]
\includegraphics[width=0.9\textwidth]{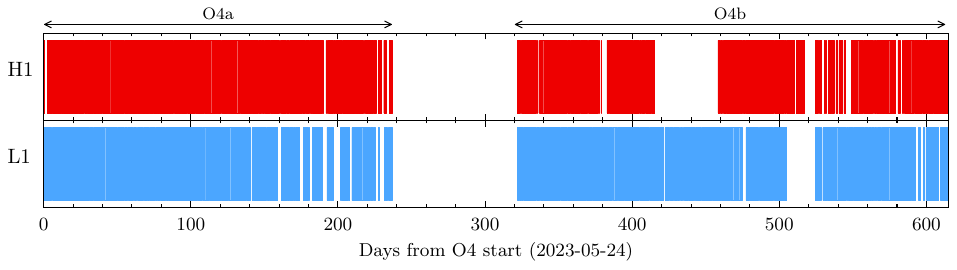}
\caption{Data availability for the LIGO Hanford (H1) and LIGO Livingston (L1) detectors during the first (O4a) and second (O4b) parts of the fourth LVK observing run~\cite{LIGOScientific:2025snk, LIGOScientific:2026jgl, LIGO:2024kkz, Capote:2024rmo}. Arrows indicate the times included in either O4a or O4b.
The long gap in H1 during O4b was due to an incident involving a damaged interferometer optic that required the detector to be taken offline from July 12 to August 24, 2024~\cite{Capote:2024rmo}.
\label{fig:sfts}}
\end{figure*}

We follow up the O4a ASAF candidates by analyzing publicly available C00-calibrated and self-gated strain data from the LIGO Hanford (H1) and LIGO Livingston (L1) detectors collected during the initial eight months of O4, known as O4a~\cite{LIGOScientific:2025snk, LIGO:2024kkz, Capote:2024rmo}.
Fig.~\ref{fig:sfts} shows the time segments analyzed in this work.
The O4a run began on May 24, 2023 at 15:00 UTC (GPS 1368975618) and ended on January 16, 2024 at 16:00 UTC (GPS 1389456018), spanning a total of $237$ days.
Even though the O4a ASAF candidates are derived from O1-O4a data, it is impractical to analyze several years of data in our follow-up due to the high computational cost and the presence of long gaps between observing runs, hence our decision to only use O4a data.
In terms of strain power spectral density (PSD), the O4a data are more sensitive over most frequencies as compared to any preceding observing run \cite{LIGO:2024kkz}.
We do not consider Virgo or KAGRA data in our analysis, as Virgo was not online during O4a and KAGRA was online for only one month.

Our pipeline ingests strain data in the form of $1800$ second Tukey-windowed short Fourier transforms (SFTs)~\cite{sftv3, lalsuite}, which are generated from the calibrated strain time series~\cite{Viets:2017yvy, Sun:2020wke, Wade:2025tgt} after applying CAT1 data quality vetoes~\cite{O4segments} and self-gating~\cite{selfgating}.
More specifically, we use the \texttt{GDS-CALIB\_STRAIN\_CLEAN\_GATED\_G02} channel from both detectors, which has undergone post-processing to remove certain known noises from the primary strain channel~\cite{LIGOScientific:2018kdd, Davis:2018yrz, Vajente:2019ycy}.
Self-gating is an additional noise-removal process designed to suppress loud noise transients (``glitches'') in the data.
For O4, a new self-gating algorithm was deployed, which iteratively determines a gating threshold by balancing noise suppression against deadtime~\cite{selfgating}.
The number of H1 SFTs in O4a is 7338 ($64.5\%$ duty factor) and the number of L1 SFTs is 7622 ($67.0\%$ duty factor).

For a select few ASAF candidates that yield outliers in our O4a search, we perform additional analysis using publicly available H1 and L1 data from O4b~\cite{LIGOScientific:2026jgl, Capote:2024rmo}, which began on April 10, 2024 at 15:00 UTC (GPS 1396796418) and ended on January 28, 2025 at 17:00 UTC (GPS 1422118818), spanning a total of 293 days.
The O4b data consist of two calibration versions: a C00 version covering the entirety of O4b, and a C01 version covering a period from September 3 to November 14, 2024, which carries the channel name \texttt{DCS-CALIB\_STRAIN\_CLEAN\_GATED\_G02}.
This C01 version was generated to correct several issues affecting the online-calibrated strain data, see Ref.~\cite{LIGOScientific:2026jgl} for details.
We use the C01-calibrated strain where available, and C00 otherwise, giving 6365 H1 SFTs in O4b ($45.2\%$ duty factor) and 9051 L1 SFTs ($64.3\%$ duty factor).

\section{Methods\label{sec:method}}

The search method employed here to follow up ASAF candidates was previously introduced in Ref.~\cite{Knee:2023toa}, which itself was built on methodology developed in previous works, e.g., Refs.~\cite{Suvorova:2016rdc, PhysRevD.97.043013, Middleton:2020skz, LIGOScientific:2021ozr, LIGOScientific:2021mwx}.
We review the follow-up pipeline here for completeness and describe its particular configuration for this work.
In addition, we describe our methods for assessing outlier significance, vetoing noise-driven outliers, and estimating search sensitivity.

\subsection{Follow-up pipeline\label{sec:search}}

The ASAF follow-up pipeline is based on the \Fstat{} matched filter augmented with hidden Markov model (HMM) frequency tracking. 
The \Fstat{} is the maximum likelihood detection statistic for CW signals in stationary Gaussian noise~\cite{Jaranowski:1998qm, Prix:2006wm, lalsuite}, and is obtained by analytically maximizing a likelihood ratio over the four amplitude parameters describing a CW signal.
The remaining phase parameters must be searched over numerically using a grid search.
For each ASAF candidate, we construct a four-dimensional grid in $(\alpha,\delta,f,\dot{f})$, where $(\alpha,\delta)$ are the source right ascension and declination (in J2000 coordinates), $f$ is the intrinsic signal frequency, and $\dot{f}$ is the frequency time derivative.
The procedure for building these parameter grids is described in Sec.~\ref{sec:grid}.
We then proceed by dividing the strain data into time segments of duration $T_\mathrm{coh}=24$~hr and evaluate the \Fstat{} at every parameter grid point for each segment.
For the O4a (O4b) dataset, this choice for the coherence time gives $N_T=237$ ($N_T=293$) time segments for the HMM.

For each ``template'' in the reduced parameter subspace $(\alpha,\delta,\dot{f})$, we employ the HMM formalism to obtain the most probable segment-wise frequency path, where the previously computed \Fstat{} values are interpreted by the HMM as emission probabilities~\cite{Suvorova:2016rdc, PhysRevD.97.043013}.
Specifically, for a fixed $(\alpha,\delta,\dot{f})$ template, we model the signal's intrinsic frequency evolution as an unbiased random walk about the (secular) frequency evolution due to the template $\dot{f}$ value, where the frequency can move at most one frequency bin, $\delta f$, from segment-to-segment.
We then apply the Viterbi algorithm to reconstruct the optimal frequency path under this constraint~\cite{Suvorova:2016rdc, PhysRevD.97.043013}.
Finally, a Viterbi log-likelihood statistic, denoted by $\Lcal$, is formed for each $(\alpha,\delta,\dot{f})$ template by incoherently summing \Fstat{} values along the frequency path.
Throughout this work, we report results in terms of the normalized log-likelihoods, $\Lbar\equiv \Lcal/N_T$, which is henceforth referred to as simply the ``detection statistic''.
Complete technical details about the implementation of HMMs and the Viterbi algorithm are provided in Refs.~\cite{Suvorova:2016rdc, PhysRevD.97.043013}.

\subsection{Parameter grids}\label{sec:grid}

Here, we describe the search ranges and step sizes of the \Fstat{} parameter grids.
For an ASAF candidate at central frequency $f_0$, the frequency grid spans the interval $f\in[f_0-\Delta f/2,f_0+\Delta f/2]$ in steps of
\begin{equation}
    \delta f = \frac{1}{2T_\mathrm{coh}}\approx 0.579\times10^{-5}~\mathrm{Hz}\,,
\end{equation} 
where $\Delta f = 1/32$~Hz, and $T_\mathrm{coh}=24$~hr is the \Fstat{} coherence time.
This gives $N_f=5401$ total frequency grid values per candidate.\footnote{
In the early stages of this work, we considered searching a broader frequency band of $\Delta f = 1/16$~Hz, i.e., doubling the search band to include half of the two adjacent ASAF bins, which would better accommodate signals that straddled the boundary between two bins.
However, we ultimately found this to be disadvantageous for candidates whose neighboring bins were contaminated with narrowband artifacts, as these artifacts could be strong enough to dominate the Viterbi maximum even when the central candidate bin is clean.
We have thus kept the search bandwidth at $\Delta f = 1/32$~Hz.
}
We define all frequencies at the midpoint of O4a, corresponding to GPS time $t_\mathrm{ref}=1379215818$ (September 20, 2023 at 3:30 UTC).

The ASAF analysis does not explicitly search over frequency derivatives.
However, an upper bound on $|\dot{f}|$ can be estimated by assuming the signal resides in the same $1/32$~Hz frequency bin throughout the entire observing period spanned by the data.
Plugging in the $8.3$~yr period spanned by O1-O4a, Ref.~\cite{LIGOScientific:2025bkz} estimates a maximum possible frequency derivative of $|\dot{f}|=1.2\times 10^{-10}$~Hz~s$^{-1}$.
Since our follow-up analysis focuses on O4a data, we assume the signal is only required to reside in one bin for the duration of O4a ($T_\mathrm{O4a}=237$~days).
This choice leads to a more conservative search interval of $\dot{f}\in[-\dot{f}_\mathrm{max},\dot{f}_\mathrm{max}]$ for each ASAF candidate, where $\dot{f}_\mathrm{max}=\Delta f/T_\mathrm{O4a}=1.526\times 10^{-9}$~Hz s$^{-1}$.
From our choice of transition matrix (Sec.~\ref{sec:search}), the maximum $\dot{f}$ trackable by the HMM is one frequency bin per coherent segment, equal to $\pm \delta f/T_\mathrm{coh}\approx \pm 6.7\times 10^{-11}$~Hz~s$^{-1}$.
To avoid saturating the $\dot{f}$ grid with many closely correlated templates, we set the $\dot{f}$ step size at
\begin{equation}
    \delta\dot{f}=2\frac{\delta f}{T_\mathrm{coh}}=\frac{1}{T_\mathrm{coh}^2}\approx 1.34\times 10^{-10}~\mathrm{Hz}~\mathrm{s}^{-1}\,,
\end{equation}
giving $N_{\dot{f}}=23$ $\dot{f}$ values per candidate. The factor of $2$ above is needed because the Viterbi algorithm can track either a positive or negative frequency drift.

Our follow-up pipeline does not explicitly search over $\ddot{f}$ or higher derivatives.
Under the ``gravitar'' model of a spinning NS with braking index $n=5$, the fastest $\ddot{f}$ consistent with the extent of our $\dot{f}$ search range is
\begin{equation}
    \ddot{f}_\mathrm{max} = \frac{n\dot{f}_\mathrm{max}^2}{f_0} \approx \frac{1.16\times 10^{-17}\:\mathrm{Hz}^2\:\mathrm{s}^{-2}}{f_0}\,.
\end{equation}
Plugging in the lowest-frequency ASAF candidate, $f_0=23.8125$~Hz, this amounts to a worst-case frequency drift of ${\sim}18$ \Fstat{} bins over O4a---small enough to be comfortably tracked by the Viterbi algorithm.

\begin{figure}[t]
\includegraphics[width=0.46\textwidth]{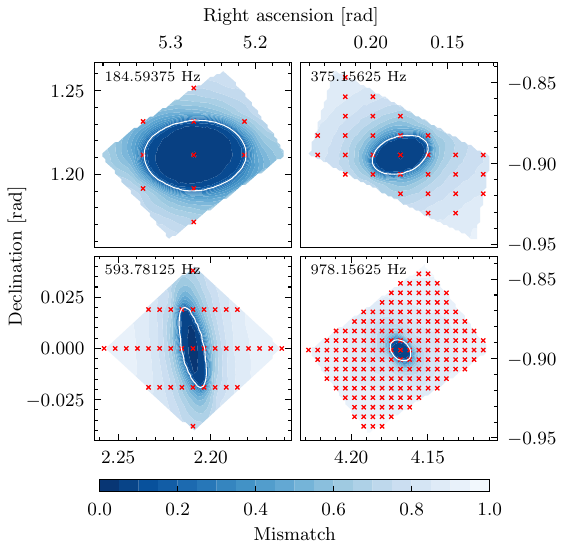}
\caption{
Mismatch contours and sky grids for a selection of four ASAF candidates at different frequencies and sky positions.
Each candidate contains a simulated signal injected into pure Gaussian noise.
The numbers in the top left corners of each panel are the injection frequencies.
The colormap shows mismatch as a function of sky position.
The white outlines are the $20\%$ mismatch contours, which are used to define the resolution of the sky grid (red crosses).
\label{fig:skygrids}}
\end{figure}

The candidate sky pixels are each tiled with rectilinear grids in the $(\alpha,\delta)$-plane, with step sizes derived empirically from injection studies.
To compute step sizes for $\alpha$ and $\delta$, we apply the following procedure to each ASAF candidate.
First, we inject a simulated signal into Gaussian noise at the central frequency and sky position of the candidate.
For simplicity, we fix the frequency derivatives of all injections to zero.
We then compute the detection statistic, $\Lbar$, on a dense grid of points covering the candidate sky pixel using the follow-up pipeline from Sec.~\ref{sec:search}.
This procedure lets us compute a signal-to-template mismatch at each point in the dense grid, defined by
\begin{equation}
    \mu = 1 - \frac{\Lbar}{\Lbar_0}\,,
\end{equation}
where $\Lbar_0$ is the detection statistic evaluated at the exact injection parameter values.
From these mismatch samples, we can extract two-dimensional contours $\mathcal{C}_{\mu^*}=\{(\alpha,\delta)\,|\,\mu(\alpha,\delta)=\mu^*\}$ corresponding to a given mismatch $\mu^*$, as shown by the examples in Fig.~\ref{fig:skygrids}.
Finally, we compute step sizes for each sky coordinate via the formula:
\begin{equation}\label{eq:deltasky}
    \delta\theta = \frac{1}{2}\bigg(\max_{(\alpha,\delta)\in\mathcal{C}_{0.2}}\theta-\min_{(\alpha,\delta)\in\mathcal{C}_{0.2}}\theta\bigg)\,, \quad \theta\in\{\alpha,\delta\}\,.
\end{equation}
Eq.~(\ref{eq:deltasky}) is equivalent to defining the step size as half the difference between the minimum and maximum values of $\alpha$ or $\delta$ lying on the $\mu^*=0.2$ mismatch contour (white outlines in Fig.~\ref{fig:skygrids}).
Tighter mismatch contours imply that signal power falls off faster as the template shifts away from the true signal parameters, requiring more densely spaced templates to compensate.

Once we have computed the required step sizes, creating the sky grid is a matter of sampling a rectangular grid centered on the candidate sky pixel, ensuring that a template always lands on the pixel origin.
A final mask is applied to the rectangular grid to discard any points that are outside the boundaries of the sky pixel (see Fig.~\ref{fig:skygrids}); we do not search on sky positions that lie outside the candidate sky pixel.

We validate our parameter grid setup by injecting simulated signals into synthetic Gaussian noise and then recovering those injections with our follow-up pipeline.
We perform $20$ injections into each candidate frequency-pixel, yielding $20\times562=11240$ total injection samples.
The injections have randomized frequencies and sky positions within the candidate frequency-pixel, and with randomized frequency derivatives within the $\dot{f}$ search range.
Since the objective of this mismatch study is to test the parameter grids, we assign loud strain amplitudes to the injections; this avoids polluting the mismatch distribution with samples that correspond to non-detections and thus have no correlation with the injected signal.
Each mismatch sample is calculated from only the max-$\Lbar$ template from the search, so that we have one mismatch sample per injection.

\begin{figure}[t]
\includegraphics[width=0.46\textwidth]{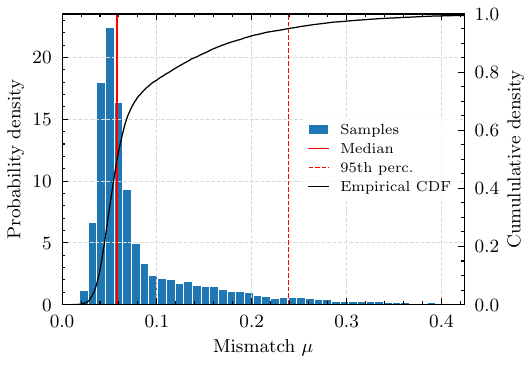}
\caption{
Mismatch distribution based on recovering simulated signals injected into ASAF frequency-pixels.
The solid and dashed vertical lines indicate the median and $95$th percentile mismatch, respectively.
The black curve is the empirical cumulative density function.
\label{fig:mismatch}}
\end{figure}

Fig.~\ref{fig:mismatch} shows the distribution of mismatches obtained from this injection study.
The distribution has a median of $\mu=0.058$ with a $95$th percentile of $\mu=0.24$, showing that our grid setup provides appropriate parameter-space coverage.
The median mismatch is smaller than in Ref.~\cite{Knee:2023toa} by roughly a factor of two, which we attribute to our doubling of the number of $f$ and $\dot{f}$ grid points (each) compared to the previous work.

\subsection{Significance}\label{sec:sig}

In order to assess the statistical significance of our results, we need to determine the probability density function of the detection statistic, $p(\Lbar)$, under the noise hypothesis.
As done in Ref.~\cite{Knee:2023toa} and other Viterbi-based CW searches, we estimate $p(\Lbar)$ by evaluating the detection statistic in real data at randomized sky positions shifted off-target from the candidate sky pixel.
Since it is assumed that no signals should exist near these off-target templates, each sample effectively behaves like a random draw from $p(\Lbar)$.
Additionally, the off-target method probes the true distribution of detector noise in the relevant frequency band, and thus is expected to provide a better approximation to $p(\Lbar)$ than simulated Gaussian noise.

We obtain off-target statistics for each ASAF candidate as follows.
We draw $(\alpha_i,\delta_i)$ samples isotropically from a band-like sky region defined by the rectangular interval $[0, 2\pi) \times [\delta_\mathrm{min}, \delta_\mathrm{max}]$, where $\delta_\mathrm{min}$ ($\delta_\mathrm{max}$) is the minimum (maximum) declination in the search template grid.
To ensure that all samples are adequately separated from the on-target search region, we accept the sample if it is at least $10$~deg from the pixel origin, and reject it otherwise.
This process is repeated until we have $N_\mathrm{off}=2000$ off-target samples.
We then evaluate the detection statistic over the full $\dot{f}$ grid at each off-target pointing, giving a total of $M=N_\mathrm{off}N_{\dot{f}}=46000$ $\Lbar_i$ samples per ASAF candidate.

Next, we model the tail of the noise distribution, $p(\Lbar)$, with an exponential function of the form
\begin{equation}\label{eq:exp}
    p(\Lbar;k,\lambda,\Ltail) = k\lambda e^{-\lambda(\Lbar - \Ltail)}\,,
\end{equation}
where $k$ is a normalization factor, $\lambda$ is a slope parameter, and $\Ltail$ is a location shift parameter.
We estimate these parameters for each ASAF candidate by fitting Eq.~(\ref{eq:exp}) to our off-target $\Lbar_i$ samples.
The location parameter $\Ltail$ can be treated as the point at which $p(\Lbar)$ becomes tail-dominated; we fix $\Ltail$ at the 96th percentile of the sample distribution, which conveniently fixes $k=0.04$ as well.
Using the remaining tail statistics, $\Lbar_i>\Ltail$, we compute the slope via the maximum likelihood estimator:
\begin{equation}
    \hat\lambda = \frac{N_\mathrm{tail}}{\sum_{i=1}^{N_\mathrm{tail}}(\Lbar_i-\Ltail)}\,,
\end{equation}
where $N_\mathrm{tail}=kM=1840$ is the number of samples in the tail.
Typical values of $\hat\lambda$ lie in the range $\hat\lambda\sim 20\text{--}25$.

Using our fitted $p(\Lbar)$ distribution, we can express the false alarm probability corresponding to a specific detection statistic value as
\begin{equation}\label{eq:pfa}
    \pfa(\Lbar) = \int_{\Lbar}^\infty p(\Lbar';k,\hat\lambda,\Ltail)\,\rmd\Lbar' = ke^{-\hat\lambda (\Lbar-\Ltail)}\,,
\end{equation}
which is simply the probability of measuring a statistic more extreme than $\Lbar$ under the noise hypothesis, i.e., a $p$-value.
An additional correction is needed to account for the trials penalty incurred by searching multiple templates for one candidate.
We thus compute the trials-factor-corrected false alarm probability for a given detection statistic as follows:
\begin{equation}\label{eq:pfacorr}
    \pcorr(\Lbar) = 1 - \big[1-\pfa(\Lbar)\big]^{N_\mathrm{tmp}}\,,
\end{equation}
where $N_\mathrm{tmp}$ is the number of $(\alpha,\delta,\dot{f})$ templates, and $\pfa(\Lbar)$ is defined in Eq.~(\ref{eq:pfa}).
One can see how searching more templates will downgrade the false alarm probability of any individual template.
Note that the calculation of $\pcorr(\Lbar)$ depends on a fit parameter ($\hat{\lambda}$) and therefore has an uncertainty.
The uncertainty propagation is described in Appendix \ref{app:pfaerror}.

In this work, we set the threshold for statistical significance at $5\%$ probability of false alarm.
We therefore consider any template satisfying $\pcorr(\Lbar)<0.05$ to be an outlier in our search.
Since $\pcorr(\Lbar)$ is a monotonically decreasing function, we can equivalently express the outlier condition as $\Lbar>\Lthres$, where $\Lthres$ is the detection statistic threshold satisfying $\pcorr(\Lthres)=0.05$.
We apply further scrutiny to any outliers through a combination of detector vetoes (Sec.~\ref{sec:veto}) and independent follow-up in O4b data (Sec.~\ref{sec:outlierfollowup}). 

\subsection{Vetoes\label{sec:veto}}

Detector noise artifacts can occasionally mimic CW signals, triggering spurious outliers in the search.
We apply three vetoes used in Ref.~\cite{Knee:2023toa} to discard outliers likely to be non-astrophysical: the known lines veto, the single interferometer veto, and the Doppler modulation (DM)-off veto.
These vetoes test whether the outlier is more consistent with originating from detector noise rather than a genuine astrophysical signal.

\begin{table*}[t]
\centering
\caption{
Parameter sampling ranges of the injections used to estimate sensitivity.
The injection frequency, $f$, and sky position, $(\alpha,\delta)$, are sampled from the frequency band and sky pixel of the ASAF candidate.
Both the frequency and phase, $\phi_0$, are defined at the midpoint of O4a, $t_\mathrm{ref}=1379215818$.
The maximum frequency derivative is $\dot{f}_\mathrm{max}=1.526\times 10^{-9}$~Hz~s$^{-1}$.
The first eight parameters are sampled for both isolated and binary injection types; the last five parameters apply to binary injections only.
The quantity $\aproj^\mathrm{max}(f,\Pbin,e)$ is defined in Eq.~(\ref{eq:aprojmax}).
Note that the $h_0$ ranges were adjusted for each candidate according to the average noise level in the candidate's frequency band; the range quoted below is the union of $h_0$ ranges from all candidates.
}
\label{tab:injections}
\begin{tabular*}{\textwidth}{@{\extracolsep{\fill}} l c c r}
\toprule
\toprule
Parameter & Symbol & Range & Distribution \\
\midrule
Frequency & $f$ & Candidate band & Uniform \\
Frequency derivative & $\dot{f}$ & $[-\dot{f}_\mathrm{max}, \dot{f}_\mathrm{max}]$ & Uniform \\
Right ascension and declination & $(\alpha,\delta)$ & Candidate sky pixel & Isotropic \\
Strain amplitude & $h_0$ & $[10^{-26},8\times 10^{-25}]$ & Uniform \\
Cosine inclination & $\cos\iota$ & $[-1,1]$ & Uniform \\
Polarization angle & $\psi$ & $[0,\pi)$ & Uniform \\
Reference phase & $\phi_0$ & $[0,2\pi)$ & Uniform \\
\midrule 
Orbital period & $\Pbin$ & $[1\:\mathrm{yr},10^3\:\mathrm{yr}]$ & Log-uniform \\
Projected semi-major axis & $\aproj$ & $[10\:\mathrm{ls},\aproj^\mathrm{max}(f,\Pbin,e)]$ & Log-uniform \\
Argument of periastron & $\omega$ & $[0, 2\pi)$ & Uniform \\
Time of periastron passage & $t_\mathrm{p}$ & $[10^9\:\mathrm{s},1.6\times 10^9\:\mathrm{s}]$ & Uniform \\
Eccentricity & $e$ & $[10^{-7},0.95]$ & Log-uniform \\
\bottomrule
\bottomrule
\end{tabular*}
\end{table*}

\subsubsection{Known lines}

The known lines veto discards outliers whose Viterbi frequency path contains any overlap with a known line artifact.
Let $f(t_n)$ denote the Viterbi frequency path (including the $\dot{f}$ contribution) of the outlier, with $1\leq n\leq N_T$.
The outlier is vetoed if the condition
\begin{equation}
    \bigg(1-\frac{v_\oplus}{c}\bigg)f_\mathrm{line}-f_\mathrm{L} \leq f(t_n) \leq  \bigg(1+\frac{v_\oplus}{c}\bigg)f_\mathrm{line}+f_\mathrm{R}\,,
\end{equation}
is true for any $n$, where $f_\mathrm{line}$ is the line frequency (typically the artifact peak), and $f_\mathrm{L,R}$ are its left and right widths. The factor of $v_\oplus/c\sim 10^{-4}$ takes into account the Doppler broadening of the line due to the Earth's orbit.
We use the list of vetted narrowband spectral artifacts observed in LIGO data during O4a~\cite{O4LIGODetector:2026okh}.
These vetted artifacts, through detector investigations and/or inspection of their spectral content, are determined to be non-astrophysical and can be safely used to veto outliers.

\subsubsection{Single interferometer}

The single interferometer veto tests whether one detector contributes a disproportionate amount of signal power, which would be indicative of a noise artifact present in one detector but not the other.
In contrast, an astrophysical signal should be present in both detectors with comparable signal strength~\cite{LIGOScientific:2017fer}.
The single interferometer veto involves searching H1 and L1 data separately with the same $(\alpha,\delta,\dot{f})$ template that produced the outlier, giving two new detection statistics $\Lbar_A\geq\Lbar_B$.
The outlier is vetoed if the larger single-detector statistic is above the original two-detector statistic,\footnote{The definition of the single interferometer veto used here differs slightly from Ref.~\cite{Knee:2023toa}. The definition in Ref.~\cite{Knee:2023toa} is more aggressive, requiring an outlier to have $\Lbar_A$ greater than the two-detector threshold, rather than the two-detector statistic, to be vetoed. In this work, we use the original definition of the single interferometer veto~\cite{LIGOScientific:2017fer} to be consistent with other Viterbi-based CW searches, e.g., Refs.~\cite{Middleton:2020skz, LIGOScientific:2021ozr}.} $\Lbar_A>\Lbar$, while the smaller single-detector statistic is sub-threshold,\footnote{
Since it is computationally expensive to repeat the off-target sampling for each single-detector search, we reuse the two-detector $\Lthres$ threshold when applying this veto.
} $\Lbar_B<\Lthres$.
In addition, we require that the frequency path associated with $\Lbar_A$ overlaps with the frequency path from the two-detector search:
\begin{equation}\label{eq:overlap}
    |f_A(t_n)-f(t_n)| \leq \frac{v_\oplus}{c}f(t_n)\,.
\end{equation}
An outlier satisfying all three of the above conditions will be vetoed, as it would be consistent with a noise artifact affecting one of the two detectors.

\subsubsection{DM-off}

The DM-off veto tests whether the signal exhibits the Doppler frequency modulations expected of an astrophysical signal~\cite{PhysRevD.106.123011}.
This veto involves repeating the search using the same template as the outlier but with Doppler timing corrections disabled inside the \Fstat{} software.
The outlier is vetoed if the DM-off search both increases the significance of the outlier, $\Lbar_\mathrm{DM-off}>\Lbar$, and recovers a similar frequency path via the overlap condition in Eq.~(\ref{eq:overlap}).
Such behavior is characteristic of a non-astrophysical artifact whose frequency is stationary in the detector rest frame.

\subsection{Sensitivity\label{sec:sensitivity}}

We estimate the sensitivity of our follow-up search by recovering simulated signals added to detector data, then fitting detection efficiency curves to the injection results.
As in Ref.~\cite{Knee:2023toa}, we report two distinct sensitivities for each ASAF candidate, corresponding either to isolated sources or sources in a long-period binary orbit with orbital period $\Pbin>1$~yr.
We consider an injection ``detected'' if a template yields an above-threshold detection statistic.

To limit computational cost, we search only $(\alpha,\delta,\dot{f})$ templates within one grid step (in each dimension) of the injection parameters.
For the same reason, we do not subject injections to the three vetoes described in Sec.~\ref{sec:veto}.
This means our sensitivities correspond to the minimum strain amplitude required for a signal to qualify as an outlier in our search before applying vetoes.
However, we still exclude candidates with disturbed backgrounds due to line contamination when reporting our results in Sec.~\ref{sec:outliers}.

\subsubsection{Signal injections}

\begin{figure}[t]
\includegraphics[width=0.45\textwidth]{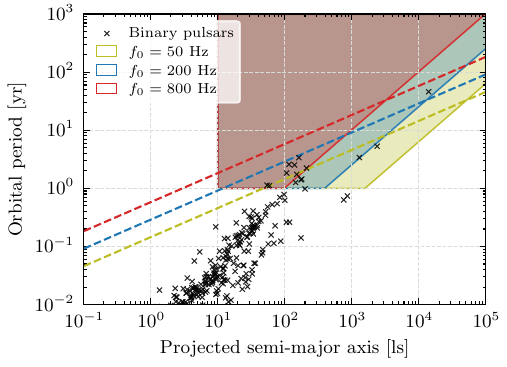}
\caption{
Binary parameter space probed by our injection recovery campaign.
The shaded regions indicate where injections are sampled in the $(\aproj,\Pbin)$-plane for different ASAF candidate frequencies.
The diagonal edge is a result of the functional bound on $\aproj$ given by Eq.~(\ref{eq:aprojmax}).
The dashed curves show where the root-mean-square Doppler shift due to (circular) orbital motion equals the maximum frequency drift allowed by the HMM. 
The black crosses show the population of known binary pulsars from the ATNF catalog~\cite{Manchester:2004bp}.
\label{fig:binaries}}
\end{figure}

For each ASAF candidate and source type, we draw a random population of $N_\mathrm{inj}=500$ test signals from the candidate sky pixel and frequency band.
The parameter sampling ranges are listed in Table \ref{tab:injections}. 
We include eccentricities up to $e=0.95$ in our binary injections.
We impose a constraint when sampling binary injections, namely that the Doppler modulation due to binary orbital motion should not be greater than the $1/32$~Hz search band.
This is done by making the upper sampling bound on the projected semi-major axis, $\aproj$, into a function of the injection parameters:
\begin{equation}\label{eq:aprojmax}
    \aproj^\mathrm{max}(f,\Pbin,e) = \min\bigg(10^5\:\mathrm{ls}, \frac{\Pbin}{2\pi} \sqrt{\frac{1-e}{1+e}}\frac{\Delta f}{2f}\bigg)\,,
\end{equation}
where $(f,\Pbin,e)$ are the injection frequency, orbital period, and eccentricity, and $\Delta f=1/32$~Hz. 
Eq.~(\ref{eq:aprojmax}) is a conservative bound corresponding to a source with orbital velocity parallel to the line-of-sight at periastron, i.e., when the Doppler effect is largest.
When sampling injections, we always sample $(f,\Pbin,e)$ first before computing the upper sampling bound on $\aproj$ via Eq.~(\ref{eq:aprojmax}).
A consequence of this functional bound is that it introduces population-level correlations between the orbital parameters, as shown in Fig.~\ref{fig:binaries}; injections at higher frequencies, shorter orbital periods, or higher eccentricities will be biased towards smaller $\aproj$.

Fig.~\ref{fig:binaries} also compares our injection parameter space with the binary pulsar population~\cite{Manchester:2004bp}.
While there are a number of binary pulsars in long-period orbits, most lie at much shorter orbital periods on the order of hours to several days. 
The decision to sample injections with $\Pbin>1$~yr is motivated by the HMM setup, which is able to track frequency derivatives of at most $|\delta f/T_\mathrm{coh}|\approx 6.7\times 10^{-11}$~Hz s$^{-1}$. 
Meanwhile, the root-mean-square (RMS) frequency derivative due to a circular binary orbit is given by
\begin{equation}\label{eq:fmaxbinary}
    \dot{f}^\mathrm{binary}_\mathrm{rms} = \frac{4\pi^2\aproj f}{\sqrt{2}\Pbin^2}\,,
\end{equation}
where $\aproj$ is measured in light-seconds, and $f$ is the signal frequency.
The Viterbi algorithm will be unable to track the signal during times when Eq.~(\ref{eq:fmaxbinary}) exceeds the maximum frequency drift allowed by the HMM.
We show where this can occur using the dashed curves in Fig.~\ref{fig:binaries}; any source below these curves has RMS frequency derivative greater than the HMM allowance at some point along its orbit.
This does not mean signals below the dashed curves cannot be detected by our search, only that they may suffer SNR loss due to high line-of-sight orbital velocity.
We also note that certain combinations of $(\aproj,\Pbin)$ can be degenerate with the template $\dot{f}$, which reduces the residual frequency drift that needs to be tracked by the HMM.
Nevertheless, Fig.~\ref{fig:binaries} shows that shorter orbital periods become increasingly difficult to track with our search algorithm, thus we only consider signals with $\Pbin>1$~yr for establishing sensitivity.

\subsubsection{Detection efficiency fitting}

We quantify sensitivity in terms of the minimum strain amplitude at which $95\%$ of signals are detected, $\hsens(\alpha,\delta,f)$, which is a function of ASAF candidate frequency and sky position.
We also report corresponding sensitivity depths, defined as
\begin{equation}\label{eq:depth}
    \depth(\alpha,\delta,f) \equiv \frac{\sqrt{S_\mathrm{n}(f)\:\mathrm{Hz}}}{\hsens(\alpha,\delta,f)}\,,
\end{equation}
where $S_\mathrm{n}(f)$ is the average noise PSD at frequency $f$.
We compute the average PSD as the harmonic mean of the PSDs from all SFTs from both detectors (i.e., not computing separate per-detector averages):
\begin{equation}
    S_\mathrm{n}(f) = \bigg(\frac{1}{N_\mathrm{SFT}} \sum_{k=1}^{N_\mathrm{SFT}} \frac{1}{S_\mathrm{n}^k(f)}\bigg)^{-1}\,,
\end{equation}
where $S_\mathrm{n}^k(f)$ is the PSD in the $k$th SFT.

We estimate $\hsens(\alpha,\delta,f)$ at every ASAF candidate using the logistic regression method.
A logistic regression fits a logistic curve to data that describes a Boolean outcome, e.g., detection or non-detection.
In our case the logistic curve represents the detection efficiency as a function of strain amplitude, and is given by
\begin{equation}\label{eq:logistic}
    \mathcal{E}(h_0;a,b)=\frac{1}{1+e^{-a (h_0-b)}}\,,
\end{equation}
where $(a,b)$ are fit parameters.
We fit Eq.~(\ref{eq:logistic}) to our injection recoveries by maximizing the likelihood function,
\begin{align}
    L(a,b) = \sum_{j=1}^{N_\mathrm{inj}}\big[&d_j\ln\mathcal{E}(h_j;a,b) \nonumber\\
    &+ (1-d_j)\ln(1-\mathcal{E}(h_j;a,b))\big]\,,
\end{align}
where $h_j$ is the strain amplitude of the $j$th injection, and $d_j$ is the detection outcome ($d_j=1$ if detected, $d_j=0$ otherwise).
We use a simple Markov chain Monte Carlo (MCMC) setup, implemented in the \textsc{emcee} Python library, to estimate the sample distributions for $a$ and $b$ given the injection data.
We then extract $\hsens$ samples by rearranging Eq.~(\ref{eq:logistic}) and substituting $\mathcal{E}=0.95$:
\begin{equation}
     \hsens = b - \frac{1}{a}\ln\bigg(\frac{1}{0.95}-1\bigg)\,.
\end{equation}
The sensitivity depth follows from Eq.~(\ref{eq:depth}).
When discussing our results in Sec.~\ref{sec:results}, we report the medians of $\hsens$ and $\depth$ for each candidate, though we note that this approach also provides uncertainties via the sample distributions.

\section{Results\label{sec:results}}

In this section, we present follow-up results for the $562$ ASAF candidates using O4a LIGO data.
We first outline our data quality-based candidate selection procedure in Sec.~\ref{sec:selection}.
We then discuss outliers found by our O4a search in Sec.~\ref{sec:outliers}.
A small subset of the ASAF candidates yield marginally significant outliers that we investigate further using LIGO data from O4b in Sec.~\ref{sec:outlierfollowup}.
We ultimately find no compelling evidence for a CW signal, concluding the discussion with sensitivity estimates in Sec.~\ref{sec:sensitivity_est}.

\subsection{Disturbed candidates\label{sec:selection}}

Although the ASAF analysis excludes certain frequency bands that contain known broadband contamination (``notch list''), these data quality vetoes do not remove all of the narrow spectral artifacts that are problematic for CW analyses.
Consequently, some of the ASAF candidates may still be contaminated.
If these artifacts are strong enough, they can severely disturb the off-target detection statistic distribution, compromising the fit to $p(\Lbar)$ and rendering the calculation of $\pcorr(\Lbar)$ unreliable.

We separate candidates with disturbed backgrounds from those with clean backgrounds by way of an empirical cut on the off-target tail threshold, $\Ltail$, which is defined as the 96th percentile of the off-target distribution per the discussion in Sec.~\ref{sec:sig}.
Spectral artifacts tend to produce inflated off-target detection statistics, which shifts the off-target distribution upwards.
This shift is captured by $\Ltail$, providing a useful proxy for identifying candidates with disturbed noise backgrounds.
We place the cut at $\Ltail>7.5$, which roughly coincides with the $95$th percentile of the $\Ltail$ values from all $562$ ASAF candidates.

\begin{figure}[t]
\includegraphics[width=0.44\textwidth]{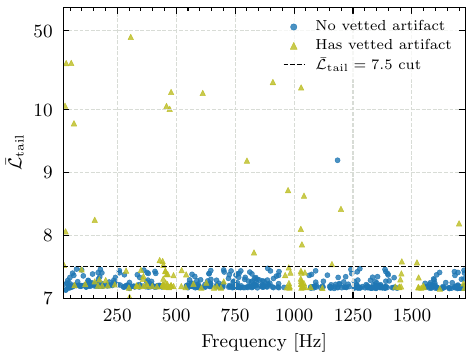}
\caption{
Detection statistic tail thresholds, $\Ltail$, for all $562$ ASAF candidates, where $\Ltail$ is defined as the 96th percentile of the off-target distribution.
The dashed horizontal line is the $\Ltail>7.5$ cut used to identify disturbed candidates.
Each candidate is also colored according to whether its band contains a vetted line artifact.
\label{fig:disturbed_bands}}
\end{figure}

The tail thresholds across all ASAF candidates are plotted in Fig.~\ref{fig:disturbed_bands}.
From the initial $562$ candidates, we identify $29$ candidates with $\Ltail>7.5$ that we subsequently mark as disturbed.
We verify the $\Ltail$ cut by cross-checking against the O4a known lines list.
Fig.~\ref{fig:disturbed_bands} shows that all but one candidate satisfying $\Ltail>7.5$ indeed contains a known vetted artifact.
The single candidate meeting this criterion that does not contain a vetted artifact instead contains an unvetted artifact in H1, i.e., a visible excursion in the H1 O4a run-averaged PSD without a known instrumental association nor a counterpart in L1 data.
We also confirmed, through visual inspection of the off-target distributions, that these $29$ candidates possess disturbed noise backgrounds.

The $29$ contaminated candidates are excluded from our main results in Sec.~\ref{sec:outliers}, as any outliers returned by these candidates would almost certainly fail to pass the vetoes\footnote{If we do not filter out disturbed candidates and take the off-target distributions at face value, we obtain two additional outliers that are both vetoed.} listed in Sec.~\ref{sec:veto}, which explicitly check for behavior consistent with spectral artifacts.
We consider the remaining $533$ ASAF candidates to have clean off-target distributions.

\subsection{Search outliers\label{sec:outliers}}

\begin{figure}[t]
\includegraphics[width=0.46\textwidth]{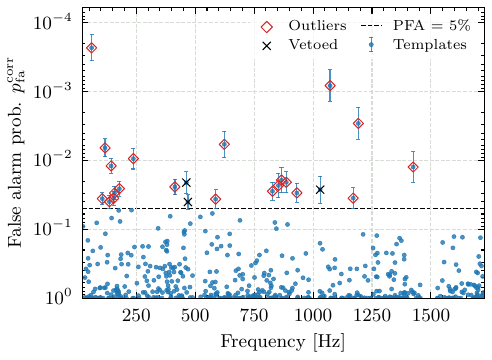}
\caption{
Follow-up results for the ASAF candidates, excluding candidates with disturbed backgrounds.
Blue points correspond to the template with the highest detection statistic obtained for each candidate, or, equivalently, the lowest false alarm probability.
Vertical error bars show the $2\sigma$ uncertainty in the estimated false alarm probabilities (shown only for the outliers to reduce clutter).
Points enclosed by a red diamond are outliers, whereas crossed-out points are vetoed.
The dashed horizontal line is the false alarm probability threshold defining an outlier, which is $5\%$ in this search.
\label{fig:outliers}}
\end{figure}

Fig.~\ref{fig:outliers} summarizes the follow-up results for the $533$ ASAF candidates passing the data quality check described in Sec.~\ref{sec:selection}.
Our analysis finds $24$ ASAF candidates that return an outlier with false alarm probability $\pcorr(\Lbar)<0.05$, after accounting for template trials factors.
The properties of the outliers are compiled in Table~\ref{tab:outliers}.
Assuming the number of outliers follows a binomial distribution, we should expect to obtain $27\pm 5$ outliers ($1\sigma$ uncertainty) from $533$ candidates using a $5\%$ false alarm probability threshold.
Therefore, finding $24$ outliers is fully consistent with statistical fluctuations.
No outliers are returned by the subset of candidates common to the O3 and O4a ASAF analyses.
The most significant outlier, at approximately $60.47$~Hz with spin-down $\dot{f}=-1.9\times 10^{-10}$~Hz~s$^{-1}$, has false alarm probability $\pcorr(\Lbar)\approx 2.3\times 10^{-4}$. 
This outlier is close to the $60$~Hz power mains fundamental, but nonetheless survives all vetoes.
The associated spin-down strain limit for this outlier is $h_0^\mathrm{sd} \sim 1.8\times 10^{-25}$, assuming a distance of $8$~kpc and a fiducial NS moment-of-inertia of $10^{38}$~kg~m$^2$, which is consistent with our estimated search sensitivities in Sec.~\ref{sec:sensitivity_est}.
We examine the outliers more closely in Sec.~\ref{sec:outlierfollowup}.

\begin{table*}[t]
\centering
\caption{
Properties of the $24$ outliers obtained by our ASAF candidate follow-up.
Columns 1--4 give the ASAF candidate parameters (frequency, sky position) and the number of templates searched for that candidate.
Columns 5--7 are the $(\alpha,\delta,\dot{f})$ parameters of the template that produced the outlier.
Column 8 gives the terminating frequency (not including the effect of $\dot{f}$) of the outlier's Viterbi path.
Columns 9--10 give the false alarm probability (with $2\sigma$ confidence intervals) of the outlier and whether or not it was vetoed, respectively.
Parameter values are rounded to the precision of the second significant digit of the grid resolution.
}
\label{tab:outliers}
\renewcommand{\arraystretch}{1.15}
\begin{tabular*}{\textwidth}{@{\extracolsep{\fill}} l c c c c c c c c r}
\toprule
\toprule
$f_0$ [Hz] & $\alpha_0$ [rad] & $\delta_0$ [rad] & $N_\mathrm{tmp}$ & $\alpha$ [rad] & $\delta$ [rad] & $\dot{f}$ [Hz s$^{-1}$] & $f(t_{N_T})$ [Hz] & $\pcorr(\Lbar)$ [\%] & Vetoed \\
\midrule
60.46875 & 5.262 & $-1.055$ & 69 & 5.34 & $-1.055$ & $-1.9\times 10^{-10}$ & 60.4753588 & 0.023$^{+0.013}_{-0.008}$ & No \\ 
105.96875 & 1.278 & 1.160 & 69 & 1.278 & 1.16 & $7.5\times 10^{-10}$ & 105.9562211 & 3.60$^{+0.78}_{-0.64}$ & No \\ 
117.46875 & 3.485 & 0.524 & 115 & 3.444 & 0.524 & $8\times 10^{-11}$ & 117.4764004 & 0.66$^{+0.23}_{-0.17}$ & No \\ 
136.03125 & 0.589 & 0.384 & 115 & 0.589 & 0.34 & $1.15\times 10^{-9}$ & 136.0249711 & 3.98$^{+0.96}_{-0.77}$ & No \\ 
143.37500 & 3.422 & $-0.840$ & 69 & 3.422 & $-0.84$ & $2.2\times 10^{-10}$ & 143.3705092 & 1.20$^{+0.34}_{-0.27}$ & No \\ 
154.12500 & 5.257 & 0.895 & 138 & 5.216 & 0.921 & $8\times 10^{-11}$ & 154.1258796 & 3.54$^{+0.91}_{-0.73}$ & No \\ 
157.90625 & 2.110 & 0.623 & 115 & 2.11 & 0.623 & $4.8\times 10^{-10}$ & 157.9201331 & 2.97$^{+0.77}_{-0.61}$ & No \\ 
178.34375 & 1.833 & 1.418 & 161 & 1.83 & 1.442 & $-9.9\times 10^{-10}$ & 178.3339988 & 2.57$^{+0.74}_{-0.58}$ & No \\ 
237.59375 & 4.432 & 0.840 & 414 & 4.48 & 0.856 & $-1.26\times 10^{-9}$ & 237.6053298 & 0.94$^{+0.39}_{-0.28}$ & No \\ 
413.78125 & 0.196 & 0.042 & 161 & 0.213 & 0.042 & $1.42\times 10^{-9}$ & 413.7800636 & 2.43$^{+0.71}_{-0.55}$ & No \\ 
462.43750 & 0.785 & $-0.675$ & 1380 & 0.755 & $-0.6661$ & $-5\times 10^{-11}$ & 462.4495312 & 2.10$^{+0.92}_{-0.64}$ & Yes \\ 
469.15625 & 3.485 & $-0.340$ & 207 & 3.471 & $-0.34$ & $1.29\times 10^{-9}$ & 469.1408275 & 4.0$^{+1.1}_{-0.9}$ & Yes \\ 
586.50000 & 5.891 & $-0.573$ & 805 & 5.9205 & $-0.572$ & $1.02\times 10^{-9}$ & 586.4979398 & 3.7$^{+1.3}_{-1.0}$ & No \\ 
623.53125 & 3.338 & 0.675 & 1955 & 3.2899 & 0.6751 & $-1.39\times 10^{-9}$ & 623.5343923 & 0.58$^{+0.32}_{-0.21}$ & No \\ 
827.65625 & 0.050 & 0.083 & 529 & 0.0417 & 0.083 & $-1.9\times 10^{-10}$ & 827.6486342 & 2.81$^{+0.99}_{-0.74}$ & No \\ 
852.93750 & 1.649 & 1.055 & 3933 & 1.6733 & 1.0547 & $-4.5\times 10^{-10}$ & 852.9310011 & 2.3$^{+1.2}_{-0.8}$ & No \\ 
866.18750 & 1.500 & 1.002 & 4278 & 1.4931 & 1.0258 & $-9.9\times 10^{-10}$ & 866.1936747 & 1.9$^{+1.0}_{-0.7}$ & No \\ 
886.40625 & 0.010 & $-0.210$ & 966 & 0.0995 & $-0.224$ & $1.29\times 10^{-9}$ & 886.4049074 & 2.08$^{+0.86}_{-0.61}$ & No \\ 
931.25000 & 3.977 & $-0.167$ & 759 & 3.9587 & $-0.167$ & $-5\times 10^{-11}$ & 931.2364178 & 3.0$^{+1.1}_{-0.8}$ & No \\ 
1030.06250 & 4.432 & 0.840 & 11730 & 4.4163 & 0.8552 & $-1.39\times 10^{-9}$ & 1030.0511863 & 2.7$^{+1.5}_{-1.0}$ & Yes \\ 
1072.25000 & 2.356 & 0.476 & 1679 & 2.3212 & 0.489 & $3.5\times 10^{-10}$ & 1072.2536516 & 0.082$^{+0.057}_{-0.034}$ & No \\ 
1170.15625 & 5.644 & $-0.430$ & 1886 & 5.6294 & $-0.419$ & $8.9\times 10^{-10}$ & 1170.1691724 & 3.5$^{+1.5}_{-1.0}$ & No \\ 
1191.84375 & 0.730 & 0.840 & 7015 & 0.7384 & 0.8252 & $-4.5\times 10^{-10}$ & 1191.8343345 & 0.29$^{+0.20}_{-0.12}$& No \\ 
1424.37500 & 2.181 & 1.107 & 21114 & 2.2608 & 1.1114 & $7.5\times 10^{-10}$ & 1424.3798379 & 1.25$^{+0.84}_{-0.50}$ & No \\ 
\bottomrule
\bottomrule
\end{tabular*}
\end{table*}

Applying the vetoes from Sec.~\ref{sec:veto} eliminates three of the $24$ outliers, leaving $21$ outliers for additional consideration.
The three vetoed outliers fail the known lines veto due to H1 artifacts:
the outliers at $462.45$~Hz and $469.14$~Hz lie in a heavily contaminated H1 band around the violin fundamental modes;
the outlier at $1030.05$~Hz is vetoed by a H1 comb artifact of unknown origin.
All three outliers are also vetoed by the single interferometer test, which correctly identifies the offending detector (H1) and confirms that the outliers behave like noise artifacts beyond simple coincidence in frequency.


\subsection{Secondary follow up\label{sec:outlierfollowup}}

We subject the $21$ non-vetoed outliers to further investigation by re-analyzing the ASAF candidates that produced the outliers using O4b LIGO data.
By following up these outliers using an independent dataset, we can assess whether they remain statistically significant in O4b and thus consistent with an astrophysical CW signal, or otherwise disappear into the noise background.

Our O4b follow-up is configured in the same manner as the initial O4a analysis.
We retain the same $(f,\dot{f})$ grid and reference time from O4a, but repeat the procedure from Sec.~\ref{sec:grid} to construct new sky grids with O4b data since the mismatch sky contours depend on the time range and duration of the data used.
We similarly repeat the off-target sampling procedure for these $21$ candidates, providing a new estimate of the $p(\Lbar)$ distribution in O4b data for evaluating false alarm probabilities.

\begin{figure}[t]
\includegraphics[width=0.44\textwidth]{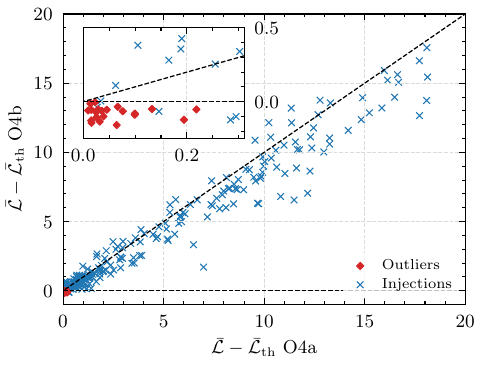}
\caption{
Follow up search results for the $21$ non-vetoed outliers (red diamonds) with O4b data.
Results for simulated signals are shown for comparison (blue crosses). 
The horizontal (vertical) axis shows the largest detection statistic minus its $5\%$ false alarm probability threshold, $\Lthres$, for the O4a (O4b) follow up.
The inset is a zoomed-in view of the lower left corner.
\label{fig:O4bfollow}}
\end{figure}

Fig.~\ref{fig:O4bfollow} shows our O4b follow-up results.
For all $21$ outliers from O4a, we do not recover a corresponding outlier with false alarm probability $\pcorr(\Lbar)<0.05$ in O4b data.
As shown in Fig.~\ref{fig:O4bfollow}, the highest detection statistics achieved in O4b are all below-threshold.
We validate this analysis by recovering simulated signals injected into both O4a and O4b data.
We perform $10$ injections per candidate, fixing their strain amplitudes at $110\%$ the estimated sensitivity $\hsens(\alpha,\delta,f)$ for the ASAF candidate, with the remaining parameters randomly sampled inside the candidate frequency-pixel.
The vast majority of the simulated signals are confidently detected in both O4a and O4b, demonstrating that an astrophysical signal that is detectable in O4a should also be detectable in O4b at similar statistical significance.

\subsection{Sensitivity estimates\label{sec:sensitivity_est}}

\begin{figure}[t]
\includegraphics[width=0.46\textwidth]{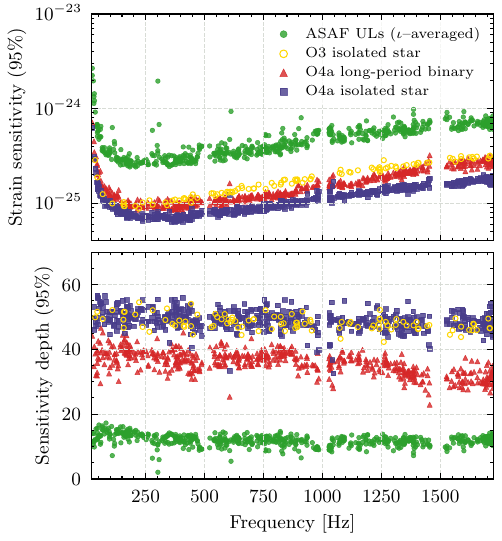}
\caption{
Estimated sensitivity of our follow up analysis at each ASAF candidate with respect to either isolated (dark blue squares) or long-period binary sources (red triangles), excluding candidates with disturbed backgrounds. 
Sensitivities from the O3 study~\cite{Knee:2023toa} (open yellow circles) and upper limits from the O4a ASAF analysis (green circles) are included for comparison~\cite{LIGOScientific:2025bkz}.
The ASAF upper limits are scaled to account for averaging over polarization.
The top panel shows sensitivity in terms of the minimum detectable strain amplitude, $\hsens(\alpha,\delta,f)$, where $(\alpha,\delta,f)$ are the candidate parameters, and the bottom panel shows sensitivity depth, $\depth(\alpha,\delta,f)$.
Error bars are omitted for visibility.
\label{fig:sensitivity}}
\end{figure}

We show sensitivity estimates for each ASAF candidate in Fig.~\ref{fig:sensitivity}, which includes sensitivities with respect to both isolated sources and sources in long-period binaries.
The sensitivities are derived from a large-scale injection campaign, as outlined in Sec.~\ref{sec:sensitivity}, and are quoted at the level of $95\%$ detection efficiency.
For comparison, we include upper limits obtained by the O4a ASAF analysis~\cite{LIGOScientific:2025bkz} alongside our own sensitivities in Fig.~\ref{fig:sensitivity}.
Since the ASAF upper limits assume a circularly polarized signal, whereas our sensitivities are averaged over polarization, we have multiplied the ASAF upper limits by an empirically derived factor of $2.3$ to account for the effect of polarization-averaging.
We also show sensitivity estimates from the previous O3 follow-up study~\cite{Knee:2023toa} in Fig.~\ref{fig:sensitivity}, illustrating the improvement in detector sensitivity between O3 and O4a.

Our analysis achieves minimum detectable strain amplitudes in the range $\hsens(\alpha,\delta,f)\sim (0.63\text{--}6.3)\times10^{-25}$ with respect to isolated sources, and $\hsens(\alpha,\delta,f)\sim (0.82\text{--}7.1)\times10^{-25}$ with respect to long-period binary sources.
The corresponding range of sensitivity depths are $\depth(\alpha,\delta,f)\sim 32.6\text{--}56.7$ for isolated sources, and $\depth(\alpha,\delta,f)\sim 22.9\text{--}46.6$ for long-period binaries.
The large sensitivity range is due to the frequency distribution of the ASAF candidates, which span the $20\text{--}1726$~Hz band, together with detector noise being highly dependent on frequency.
We are more sensitive to isolated sources than long-period binary sources since the latter tend to exhibit more rapid frequency evolution from the orbital Doppler shift.
The sensitivity of our follow-up pushes well below the polarization-averaged upper limits set by the ASAF analysis by a factor of ${\sim}4.1$ in strain amplitude for isolated sources.
If we ignore the inconsistent polarization conventions and simply compare our polarization-averaged sensitivities to the circularly polarized ASAF upper limits, the improvement factor becomes ${\sim}1.8$. 
Comparing our O4a isolated-star strain sensitivities with O3~\cite{Knee:2023toa}, we find an average improvement factor of ${\sim}1.6$.
We achieve our best sensitivity in the $100\text{--}300$~Hz band, where our results are competitive with the most stringent all-sky upper limits obtained using highly model-dependent templated searches in O4a data~\cite{LIGOScientific:2026plm}.

\section{Conclusions\label{sec:conclusions}}

We have presented a follow-up search for CW signals triggered on $562$ sub-threshold candidates identified by the O4a ASAF analysis~\cite{LIGOScientific:2025bkz}.
We performed the primary follow-up in O4a LIGO data, using O4b data for secondary follow-up of statistically significant outliers.
The search pipeline combines the \Fstat{} matched filter with HMM frequency tracking, accommodating signals with frequency evolution that deviates from the canonical phase model of an isolated spinning NS.
After filtering out candidates with disturbed frequency bands, we found $24$ ASAF candidates yielding an outlier with false alarm probability below $5\%$ after correcting for trials factors.
Considering the total number of candidates analyzed, all outliers are of relatively marginal significance and the number of outliers is consistent with our chosen false alarm probability threshold.
Three outliers resemble detector noise artifacts and were vetoed outright.
We subjected the remaining $21$ outliers to independent follow-up using O4b LIGO data.
None of the outliers persist into O4b, and we therefore conclude that all outliers are consistent with noise fluctuations in O4a data.
The outlier parameters are recorded in Table \ref{tab:outliers} for future reference.

Through a large-scale injection campaign, we estimate the sensitivity of our search at each ASAF frequency-pixel with respect to either isolated or long-period binary sources.
The peak polarization-averaged sensitivity achieved at any single ASAF candidate is $\hsens(\alpha,\delta,f)=6.3\times 10^{-26}$ for isolated sources and $\hsens(\alpha,\delta,f)=8.2\times 10^{-26}$ for long-period binary (with $\Pbin>1$~yr) sources at frequencies of $373.8$~Hz and $409.9$~Hz, respectively.

Despite the null result, this work highlights the importance of following up candidates derived from stochastic GW search pipelines.
Thanks to its flexible, model-agnostic paradigm, the ASAF analysis can detect quasi-monochromatic signals that do not follow the fully phase-coherent signal model assumed in most CW searches.
Yet, there are numerous reasons to believe that some CW sources are not persistently coherent, such as NSs that are actively accreting or that undergo frequency glitching behavior. 
Hierarchical follow-up of stochastic candidates using CW search techniques fills this critical blind spot present in CW all-sky searches, thus expanding the parameter space of detectable signals.
This work demonstrates that our ASAF follow-up search pipeline has reached a mature enough stage that it can be deployed at-scale to analyze hundreds of candidates.

\appendix

\section{False alarm probability uncertainty\label{app:pfaerror}}

In Sec.~\ref{sec:outliers}, we presented our search results in terms of the false alarm probability, $\pcorr(\Lbar)$, defined in Eqs.~\ref{eq:pfa} and \ref{eq:pfacorr}.
Because $\pcorr(\Lbar)$ is derived from a maximum likelihood fit to the noise distribution (see Eq.~(\ref{eq:exp})), the uncertainty in the slope parameter, $\hat{\lambda}$, will propagate into the final estimate of $\pcorr(\Lbar)$.
For completeness, we describe the uncertainty propagation here.

In the Gaussian approximation, the standard error of $\hat{\lambda}$ is
\begin{equation}
    \sigma_{\hat{\lambda}} = \frac{\hat{\lambda}}{\sqrt{N_\mathrm{tail}}}\,,
\end{equation}
where we recall that $N_\mathrm{tail}$ is the number of off-target samples used to estimate $\hat{\lambda}$.
Since probabilities cannot be negative, it is convenient to work in terms of log-probabilities,
\begin{equation}
    \ln\pfa(\Lbar)=\ln k -\hat{\lambda}(\Lbar-\Ltail)\,.
\end{equation}
Using the basic propagation rules, the uncertainty in $\ln\pfa(\Lbar)$ is
\begin{equation}
    \sigma_{\ln\pfa} = \sigma_{\hat{\lambda}}(\Lbar-\Ltail)\,.
\end{equation}
As expected, the error grows linearly with $\Lbar$, which is a result of extrapolating the exponential fit farther out into the distribution tail.
We then compute asymmetric $2\sigma$ errors on $\pfa(\Lbar)$ as
\begin{equation}
    \pfa^\pm(\Lbar) = e^{\ln\pfa(\Lbar) \pm 2\sigma_{\hat{\lambda}}(\Lbar-\Ltail)}\,.
\end{equation}
Finally, applying the trials factor correction from Eq.~(\ref{eq:pfacorr}) to each of $\pfa^-(\Lbar)$ and $\pfa^+(\Lbar)$ yields the $2\sigma$ confidence interval in $\pcorr(\Lbar)$.
These intervals are shown as error bars in Fig.~\ref{fig:outliers} and Table \ref{tab:outliers}.

\section{Computing resources}

Our analysis workflow leverages the Directed Acyclic Graph (DAG) framework implemented in HTCondor~\cite{htcondor}.
Each DAG manages one ASAF candidate, including the on-target follow-up, off-target sampling, injection recovery, sensitivity estimation, and diagnostic plot generation.
The workload is distributed such that each HTCondor job processes all $\dot{f}$ templates for a single template sky position $(\alpha,\delta)$.
The pipeline runs entirely on CPUs.
We performed the bulk of our analysis on distributed computing clusters supported by the Open Science Grid (OSG)~\cite{Pordes_2007, 5171374, osg1, osg2}.

Our full O4a analysis processed a total of $2{,}813{,}659$ on-target $(\alpha,\delta,\dot{f})$ templates and $25{,}852{,}000$ off-target templates across all ASAF candidates, consuming ${\sim}1.2\times 10^6$ CPU core-hours.
Most of this processing time is spent on gathering off-target statistics.
Candidates at higher frequencies involve more on-target templates and thus require more processing time than lower-frequency candidates.

\section{Validation with hardware injections\label{app:hwinj}}

We validate the performance of our follow up pipeline by attempting to recover pulsar hardware injections.
Instead of adding a waveform to data in software, hardware injections are created by pushing on the interferometer test masses using a secondary laser system (the photon calibrator), thereby achieving the equivalent physical effect that a real GW would have on the detector~\cite{Karki:2016pht}.
Throughout O4, the LVK CW group maintained a set of continuously running CW hardware injections for the purposes of monitoring detector performance and validating search pipelines~\cite{Baxi:2026wbr}.
We test our pipeline on five of these hardware injections using O4a data only; their parameters are listed in Table \ref{tab:hwinj}.
We choose these five as they are well separated in frequency and sample the low-, mid-, and high-frequency range of the detector band.

\begin{table}[t]
\centering
\caption{
Parameters of the five pulsar hardware injections recovered with our search pipeline.
Frequencies are quoted at the midpoint time of O4a. 
}
\label{tab:hwinj}
\begin{tabular*}{\columnwidth}{@{\extracolsep{\fill}} l c c c r}
\toprule
\toprule
Inj & $f(t_\mathrm{ref})$ [Hz] & $\dot{f}$ [Hz s$^{-1}$] & $h_0$ & Detected \\
\midrule
0 & $265.5745009$ & $-4.15\times 10^{-12}$ & $4.01\times 10^{-26}$ & Yes \\
1 & $848.8950355$ & $-3\times 10^{-10}$ & $2.77\times 10^{-25}$ & Yes \\
2 & $575.163487$ & $-1.37\times 10^{-13}$ & $3.2\times 10^{-26}$ & Yes \\
3 & $108.8571594$ & $-1.46\times 10^{-17}$ & $1.3\times 10^{-25}$ & Yes \\
4 & $1387.2239216$ & $-2.54\times 10^{-8}$ & $4.9\times 10^{-25}$ & No \\
\bottomrule
\bottomrule
\end{tabular*}
\end{table}

We configure our pipeline identically to the O4a ASAF follow up analysis,
with the same choice of coherence time, search bandwidth, $\dot{f}$ grids, sky pixel basis, and $5\%$ false alarm probability threshold.
Since the \textsc{HEALPix}~\cite{Gorski:2004by, Zonca2019} basis has fixed pixel centers, we simply select the sky pixel containing the target injection.

\begin{figure}[t]
\includegraphics[width=0.42\textwidth]{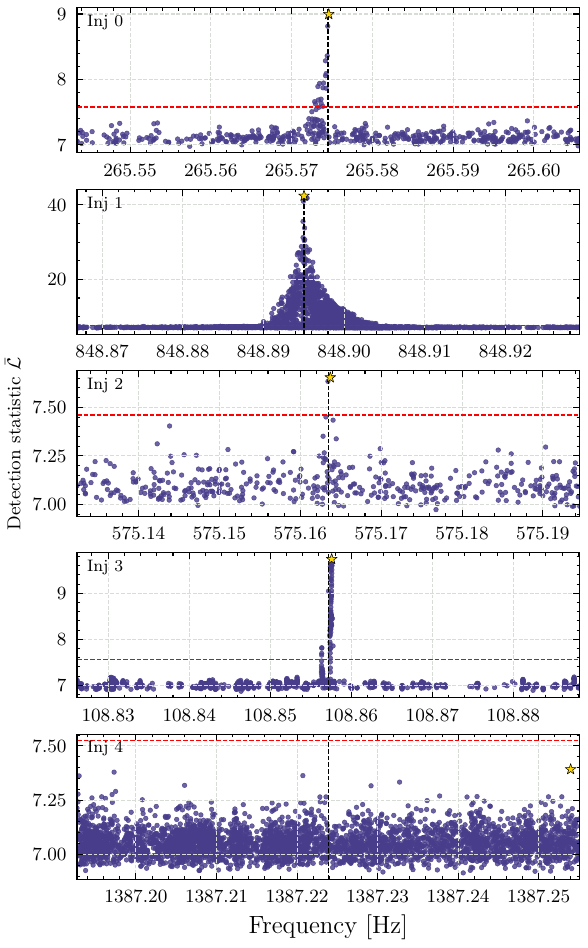}
\caption{
Search results for five selected pulsar hardware injections, using the same search pipeline for following up ASAF candidates.
Each point corresponds to a single $(\alpha,\delta,\dot{f})$ template.
Vertical dashed lines show the frequency of the injection at the reference time of the search, and the horizontal dashed lines show the threshold detection statistic corresponding to $\pcorr(\Lthres)=0.05$.
Star markers indicate the maximum-$\Lbar$ template for each injection.
\label{fig:hwinj}}
\end{figure}

Fig.~\ref{fig:hwinj} shows the detection statistic distribution for each hardware injection.
Our pipeline detects four of the five hardware injections.
The missed injection, Inj 4, possesses a very fast spin-down that is well outside the $\dot{f}$ search range. 

\begin{acknowledgments}

We thank Jishnu Suresh, Deepali Agarwal, and the LVK Continuous Waves group for helpful discussions. 
We thank Graham Woan for providing comments on this manuscript.
Some of the results in this paper have been derived using the \textsc{HEALPix}~\cite{Gorski:2004by, Zonca2019} package.
This material is based upon work supported by NSF’s LIGO Laboratory which is a major facility fully funded by the National Science Foundation.
The authors are grateful for computational resources provided by the LIGO Laboratory through National Science Foundation Grants PHY-0757058 and PHY-0823459.
This research was done using services provided by the Open Science Grid Consortium~\cite{Pordes_2007, 5171374, osg1, osg2}, which is supported by the National Science Foundation Awards OAC-2030508 and OAC-2323298.
A.M.K. and K.R. acknowledge funding support from the National Science Foundation Award PHY-2408883.
L.S. acknowledges the support by the Australian Research Council (ARC) Centre of Excellence for Gravitational Wave Discovery (OzGrav), Project Number CE230100016, and the ARC Discovery Early Career Researcher Award, Project Number DE240100206. 

\end{acknowledgments}

\bibliography{biblio}

\end{document}